\documentclass[12pt]{iopart}
\usepackage{graphicx}

\begin{document}

\title[Particle ratios and identified hadron spectra from PHOBOS]
{Antiparticle to particle ratios and identified hadron spectra 
in Cu+Cu and Au+Au collisions}

\author{G\'abor I Veres for the PHOBOS Collaboration\\
%
%
\vspace{3mm}
\footnotesize
B.Alver$^4$,
B.B.Back$^1$,
M.D.Baker$^2$,
M.Ballintijn$^4$,
D.S.Barton$^2$,
R.R.Betts$^6$,
A.A.Bickley$^7$,
R.Bindel$^7$,
W.Busza$^4$,
A.Carroll$^2$,
Z.Chai$^2$,
V.Chetluru$^6$,
M.P.Decowski$^4$,
E.Garc\'{\i}a$^6$,
N.George$^2$,
T.Gburek$^3$,
K.Gulbrandsen$^4$,
C.Halliwell$^6$,
J.Hamblen$^8$,
I.Harnarine$^6$,
M.Hauer$^2$,
C.Henderson$^4$,
D.J.Hofman$^6$,
R.S.Hollis$^6$,
R.Ho\l y\'{n}ski$^3$,
B.Holzman$^2$,
A.Iordanova$^6$,
E.Johnson$^8$,
J.L.Kane$^4$,
N.Khan$^8$,
P.Kulinich$^4$,
C.M.Kuo$^5$,
W.Li$^4$,
W.T.Lin$^5$,
C.Loizides$^4$,
S.Manly$^8$,
A.C.Mignerey$^7$,
R.Nouicer$^2$,
A.Olszewski$^3$,
R.Pak$^2$,
C.Reed$^4$,
E.Richardson$^7$,
C.Roland$^4$,
G.Roland$^4$,
J.Sagerer$^6$,
H.Seals$^2$,
I.Sedykh$^2$,
C.E.Smith$^6$,
M.A.Stankiewicz$^2$,
P.Steinberg$^2$,
G.S.F.Stephans$^4$,
A.Sukhanov$^2$,
A.Szostak$^2$,
M.B.Tonjes$^7$,
A.Trzupek$^3$,
C.Vale$^4$,
G.J.van~Nieuwenhuizen$^4$,
S.S.Vaurynovich$^4$,
R.Verdier$^4$,
G.I.Veres$^4$,
P.Walters$^8$,
E.Wenger$^4$,
D.Willhelm$^7$,
F.L.H.Wolfs$^8$,
B.Wosiek$^3$,
K.Wo\'{z}niak$^3$,
S.Wyngaardt$^2$,
B.Wys\l ouch$^4$\\
%
\vspace{3mm}
\scriptsize
$^1$~Argonne National Laboratory, Argonne, IL 60439-4843, USA\\
$^2$~Brookhaven National Laboratory, Upton, NY 11973-5000, USA\\
$^3$~Institute of Nuclear Physics PAN, Krak\'{o}w, Poland\\
$^4$~Massachusetts Institute of Technology, Cambridge, MA 02139-4307, USA\\
$^5$~National Central University, Chung-Li, Taiwan\\
$^6$~University of Illinois at Chicago, Chicago, IL 60607-7059, USA\\
$^7$~University of Maryland, College Park, MD 20742, USA\\
$^8$~University of Rochester, Rochester, NY 14627, USA\\}

\ead{\mailto{veres@mit.edu}}

\begin{abstract}
New results on antiparticle to particle ratios in Cu+Cu and Au+Au 
collisions at $\sqrt{s_{_{NN}}}$ = 62.4 and 200 GeV from the PHOBOS 
experiment at RHIC are presented.
Transverse momentum spectra of pions, kaons, protons and antiprotons
from Au+Au collisions at $\sqrt{s_{_{NN}}}$ = 62.4 GeV close to 
mid-rapidity are also discussed.
Antiparticle to particle ratios are found to be remarkably independent of 
the collision centrality in both colliding systems. The collision energy 
dependence of the $\overline{p}/p$ ratios is very significant in Cu+Cu 
collisions. Baryons are found to have substantially harder
transverse momentum spectra than mesons.
The $p_T$ region in which the proton to pion ratio reaches unity in
central Au+Au collisions at $\sqrt{s_{_{NN}}}$ = 62.4 GeV
fits into a smooth trend as a function of collision energy.
The observed particle yields at very low $p_T$ are comparable to
extrapolations from higher $p_T$ for kaons, protons and antiprotons.
The net proton yield at mid-rapidity is found to be proportional 
to the number of participant nucleons in Au+Au collisions at 62.4 and 
200 GeV energies.
\end{abstract}
\submitto{\JPG}
\section{Introduction}

The antiparticle to particle ratios of hadrons are important 
observables of heavy ion collisions, and have been extensively studied
since the very beginning of RHIC operations
\cite{phobosrat:aa130,starrat:aa130}. They are more precisely 
measurable than invariant cross sections, and provide quantitative input 
to models concentrating on the hadro-chemistry of the collisions, such as 
statistical thermal models. Antiproton to proton ratios and net 
proton yields provide 
information on baryon transport and baryon production in these 
ultra-relativistic collisions, and are relevant quantities to estimate the 
fraction of the total collision energy that is available for particle 
production \cite{Bearden:2003hx}.

The presented identified hadron spectra at 
$\sqrt{s_{_{NN}}}$ = 62.4 GeV extend the 
energy range for which the contributions of different particle species to 
the inclusive charged hadron spectra are known, bridging a gap between 
17.2 and 130 GeV. Invariant cross sections at very low $p_T$, net 
proton yields and baryon enhancement in the intermediate $p_T$ range 
(reported earlier in \cite{Adler:2003cb,Adcox:2001mf,Adcox:2003nr}) are 
studied in Au+Au collisions at 62.4 GeV. 


The sensors in the two arms of the PHOBOS silicon spectrometer 
are arranged into layers \cite{phobos:si}. 
The particles with $p_T$ measured to be less than 200 MeV/c range 
out in these layers. Particle ratios are measured in the $0.2<p_T<0.8$ 
GeV/c range in the spectrometer, and the identified spectra analysis uses
the time-of-flight detector as well, extending the $p_T$ reach to 3 
GeV/c.

The acceptance corrections of the particle ratios are cancelled by 
changing the polarity of the magnetic field: the ratios are formed from 
the yields of the positive and negative particles with similar 
trajectories corresponding to the same bending direction, but to opposite 
magnetic polarity.
Ratios in each bending direction and spectrometer arm are evaluated 
separately and averaged, at the same time providing a handle on the 
systematic errors. The ratios are corrected for absorption in the beam-pipe 
and in the detector materials, secondary particle production and feed-down 
from weak decays, based on detailed Monte Carlo simulations.
The size of the latter correction 
is only 1-2\%, since the first silicon layer is positioned within 10~cm 
from the interaction point, thus allowing good rejection of tracks from 
displaced vertices. 

\section{Results}

\begin{figure}[t]
\centerline{\includegraphics[width=54.5mm]{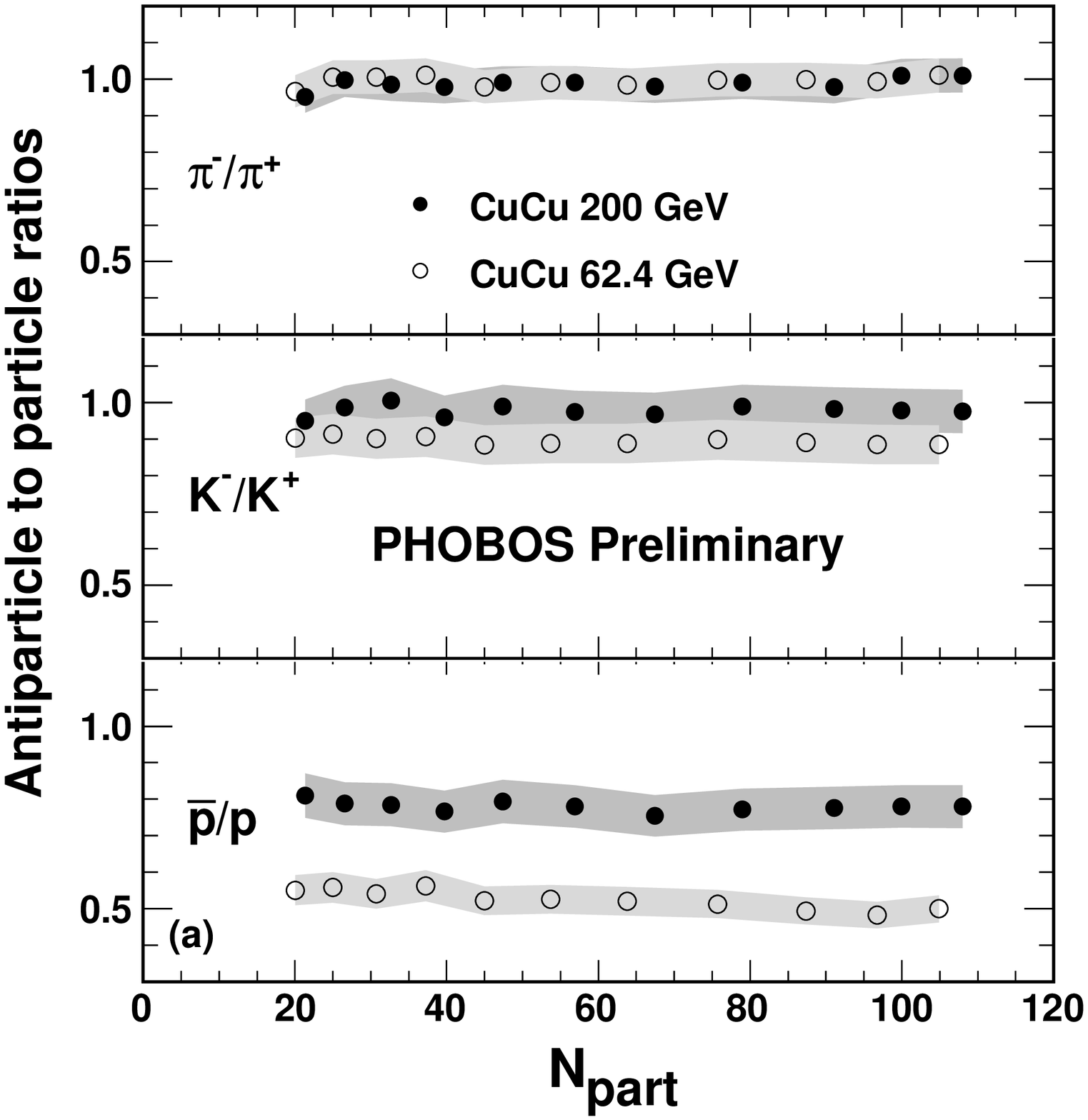}
\includegraphics[width=54.5mm]{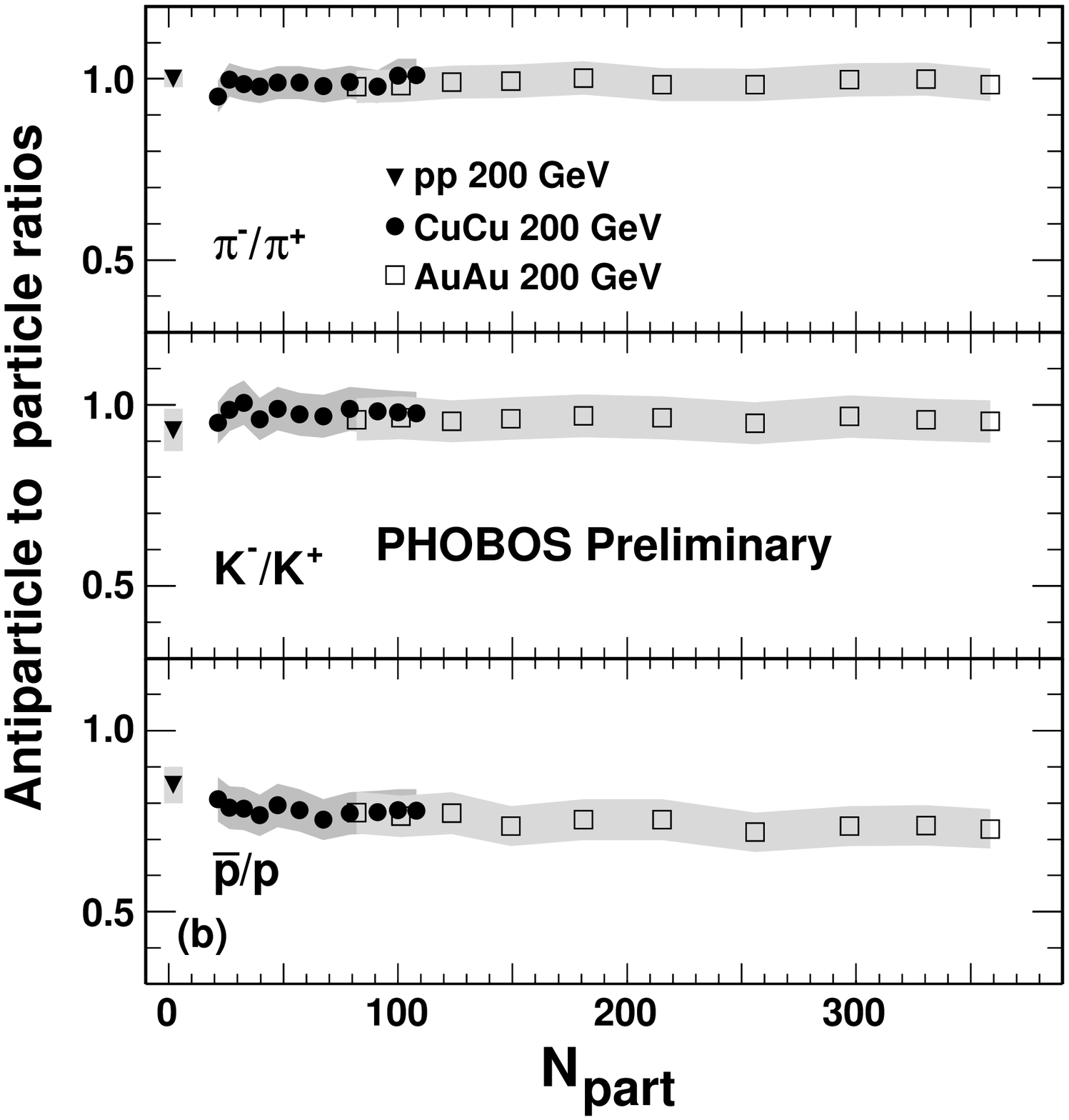}}
\caption{\label{ratios}
Preliminary $\pi^-/\pi^+$ (top), $K^-/K^+$ (center) and $\overline{p}/p$ 
(bottom) ratios as a function of the number of participant nucleons. 
(a) Cu+Cu collisions at $\sqrt{s_{_{NN}}}$ = 62.4 GeV (open points)   
and 200 GeV (closed symbols). (b) Comparison of p+p (triangles),
Cu+Cu (dots) and Au+Au (open squares) collisions at $\sqrt{s_{_{NN}}}$ =
200 GeV. The gray bands correspond to systematic errors at 90\% C.L.}
\end{figure}

The preliminary $\pi^-/\pi^+$, $K^-/K^+$ and 
$\overline{p}/p$ ratios in p+p, Cu+Cu and Au+Au collisions are presented 
in figure \ref{ratios} as a function of the number of participants in the 
collision \cite{Phobos:pid62}. 
The comparison of different collision energies (left panel) 
shows that the $\overline{p}/p$ ratios are strongly energy dependent in 
Cu+Cu collisions (similarly to collisions of heavier ions). 
There is only an insignificantly small, if any, decrease of the 
$\overline{p}/p$ ratio observed with increasing collision centrality
at both energies. On the right panel, the same ratios in p+p, Cu+Cu and 
Au+Au collisions are compared at 200 GeV, where again, almost no 
variation of the $\overline{p}/p$ ratio with centrality is visible. This 
observation challenges theoretical expectations about 
baryon stopping in collisions with varying impact parameter.
These comparisons are being extended to 62.4 GeV where the ratios are 
further away from unity, possibly leaving more room for 
measurable variations.

\begin{figure}[t] 
\centerline{\includegraphics[width=71mm]{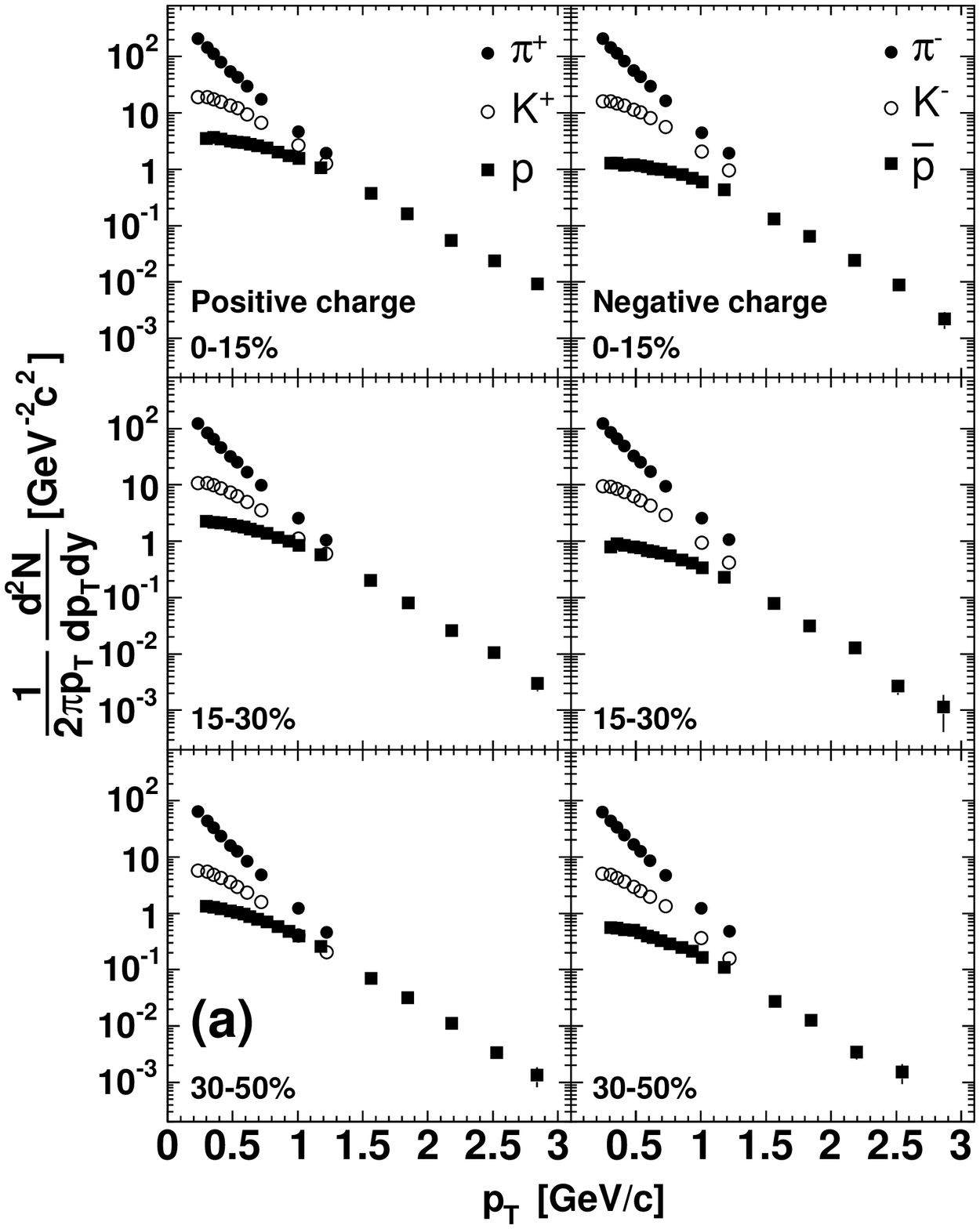}
\includegraphics[width=60mm]{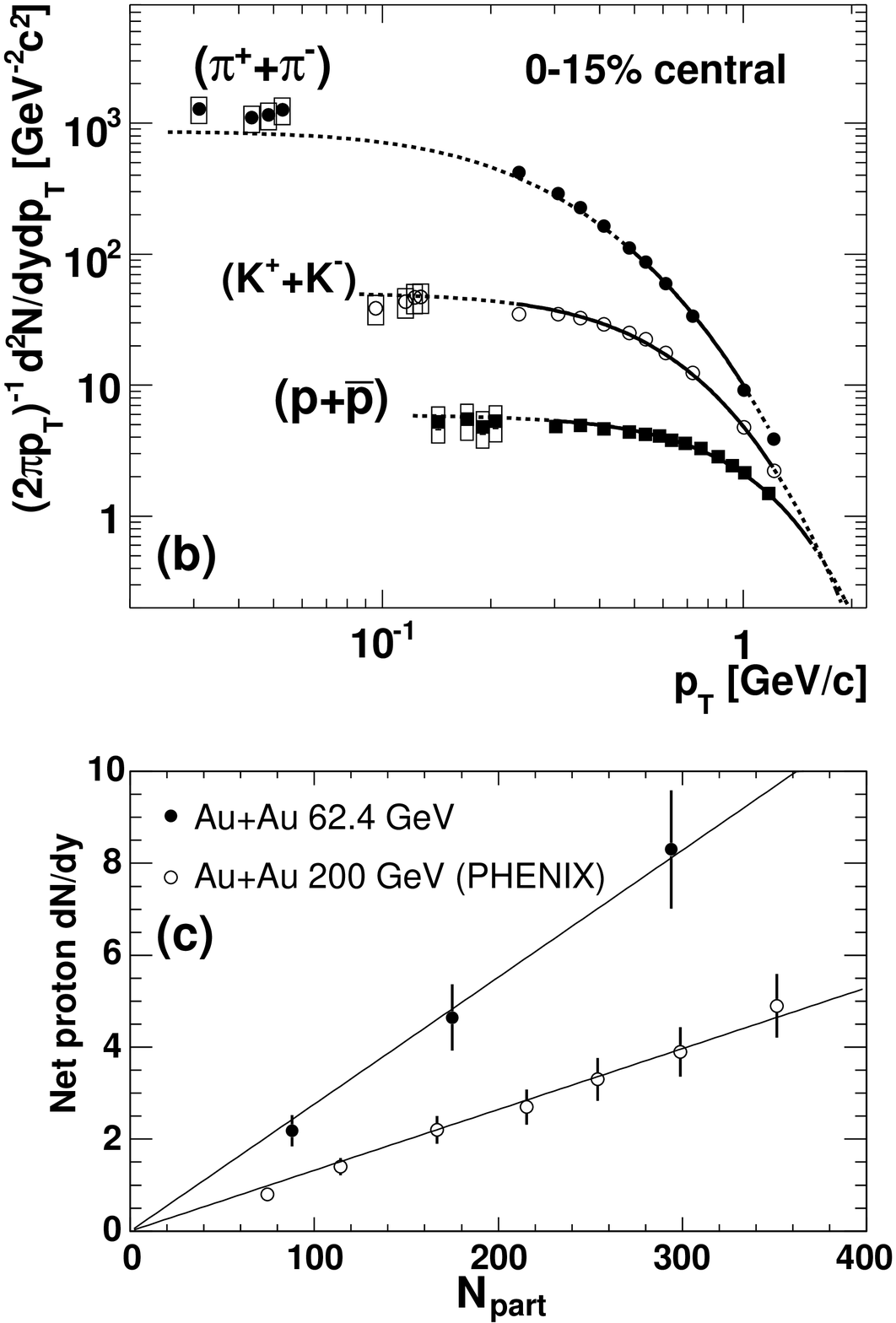}} 
\caption{\label{spectra}
(a) Invariant cross sections of identified hadrons in Au+Au collisions at
$\sqrt{s_{_{NN}}}$ = 62.4 GeV. Positives are
plotted in the left, negatives in the right~column.
(b) Invariant cross sections of $\pi^++\pi^-$, $K^++K^-$ and
$\overline{p}+p$, including~the very low $p_T$ data in the most central class.
(c) Net proton yield close to mid-rapidity as a function of $N_{part}$ in
Au+Au collisions at $\sqrt{s_{_{NN}}}$ = 62.4 and 200 GeV.}
\end{figure}

In order to be able to study the ratios of different particle yields or
$p_T$ integrated $dN/dy$ rapidity densities, one has to measure invariant 
spectra of identified particles as well. Figure 
\ref{spectra}a summarizes the measured $p_T$ spectra of charged hadrons
in Au+Au collisions at $\sqrt{s_{_{NN}}}$ = 62.4 GeV. 
Different rows correspond to three classes of centrality 
(expressed as the percentile of the total inelastic Au+Au cross section, 
starting with the 15\% most central collisions in the top row). 
A smooth evolution of the spectra with centrality can be observed, 
with the proton spectrum being harder than the meson spectra, similar to 
results at higher collision energies \cite{Adler:2003cb,Adams:2003xp}.
Figure \ref{spectra}b compares the $\pi^++\pi^-$, $K^++K^-$ and
$\overline{p}+p$ spectra to the cross sections measured at very low $p_T$.
A blast wave fit is applied to the higher $p_T$ data points (solid lines),
and extrapolated to low $p_T$ (dashed lines). The extrapolation is 
in good agreement with the kaon and (anti)proton yields (the fit does not 
take into account resonance production which populates the low $p_T$ 
region with pions).

The proton and antiproton spectra can be integrated over $p_T$, with only 
a small extrapolation of the yield to the unmeasured low and high-$p_T$ 
region. Figure \ref{spectra}c shows the integrated 
yield, $dN/dy$, of the net protons ($\overline{p}-p$) close to 
mid-rapidity at 62.4 \cite{Phobos:pid62} and 200 GeV \cite{Adler:2003cb} 
collision energies. In both cases, the net 
proton yield is approximately proportional to $N_{part}$, which does not 
meet expectations of increasing amount of baryon stopping with increasing 
centrality. According to figure~\ref{ratios}, the $\overline{p}/p$ ratio 
does not depend strongly on centrality, thus the $\overline{p}$ and $p$ 
yields are also proportional to $N_{part}$ to a good approximation.

\begin{figure}[t]
\centerline{\includegraphics[width=65mm]{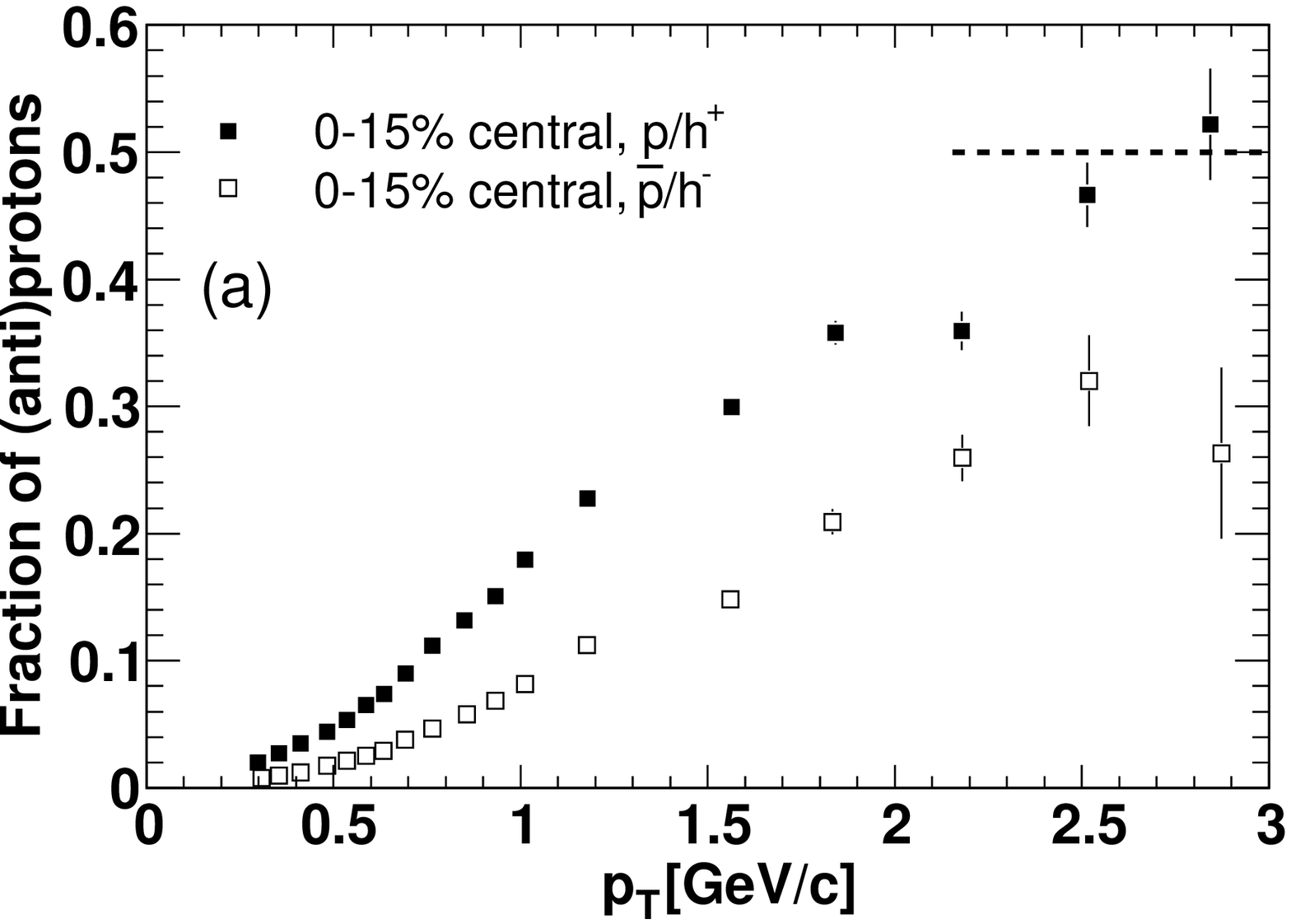}
\includegraphics[width=63mm]{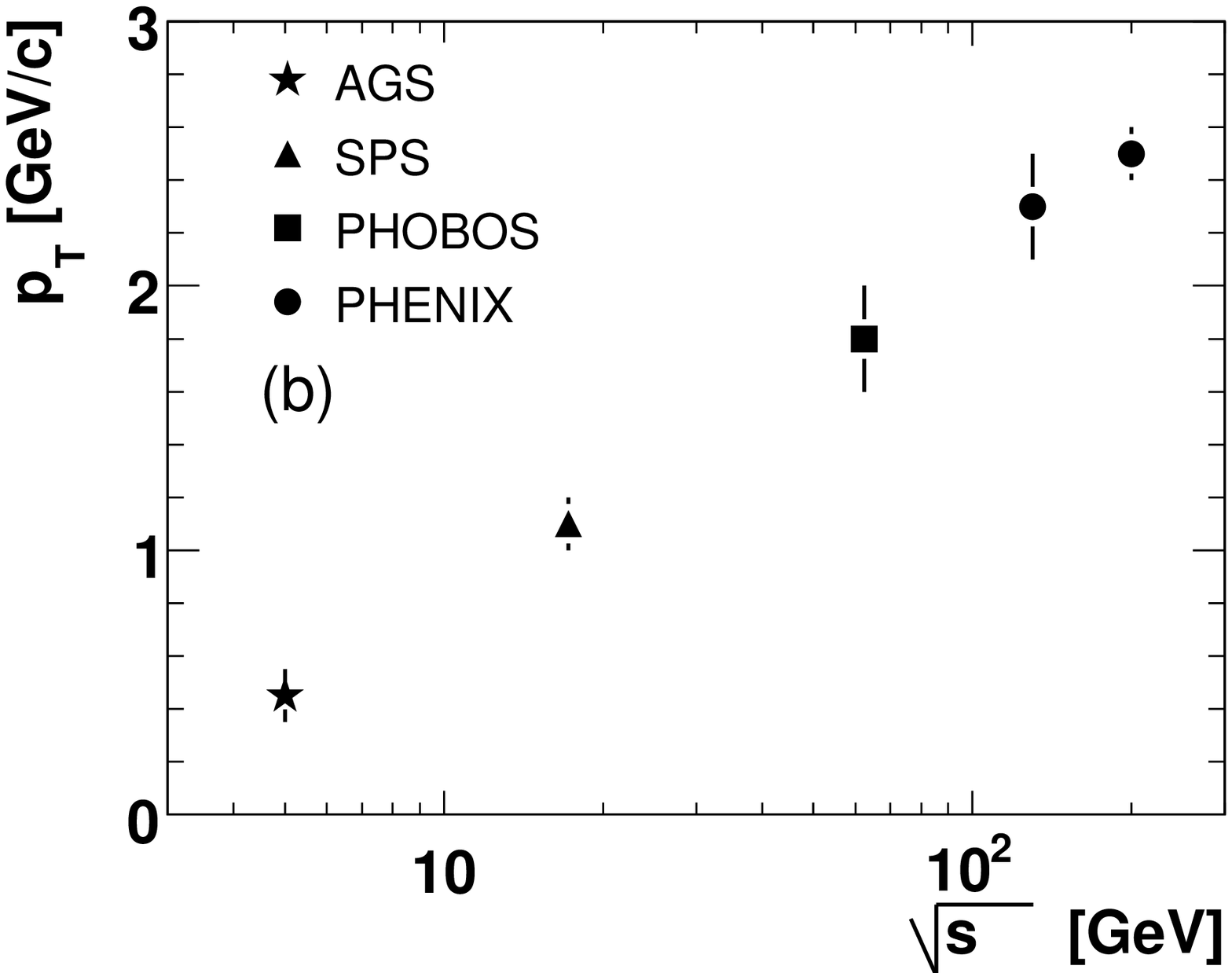}} 
\caption{\label{fractions}
(a) $p/h^+$ (closed symbols) and $\overline{p}/h^-$ (open
symbols) ratios as a function of $p_T$ in Au+Au collisions
at $\sqrt{s_{_{NN}}}$ = 62.4 GeV. (b) The $p_T$ value where the
proton and $\pi^+$ invariant yields become equal in central Au+Au (Pb+Pb)
collisions, as a function of $\sqrt{s_{_{NN}}}$. All data have been
corrected for feed-down from weak decays (see text for references).}
\end{figure}

The less steeply falling proton $p_T$ spectrum dominates over the mesons 
at higher $p_T$ values, and at 200 GeV collision energy, even 
the antiproton yield can exceed the pion yield at high $p_T$ 
\cite{Adcox:2001mf}. The phenomenon may be important to identify the relevant 
degrees of freedom and to study the energy loss of partons
in the medium created in a heavy ion 
collision. To illustrate the baryon dominance at high 
$p_T$, the fraction of protons among all positive hadrons ($p/h^+$)
and the fraction of antiprotons among all negative hadrons 
($\overline{p}/h^-$) are plotted in figure 
\ref{fractions}a, for central Au+Au collisions at 62.4 GeV.
The $p/h^+$ ratio approaches 1/2 around 2.5--3 GeV/c (dashed line), which 
indicates that the proton yield becomes dominant over both the $\pi^+$ 
and $K^+$ yields.

Figure~\ref{fractions}b illustrates the evolution of 
the `baryon dominance' with collision energy.
For central heavy-ion (Au+Au or Pb+Pb) collisions at the AGS, SPS and RHIC 
accelerators, the approximate $p_T$ value at which the invariant cross 
sections of protons and positive pions at mid-rapidity approach each 
other is plotted.
Data (with the percentile of most central events) are taken 
from the 
E802~\cite{Ahle:1998jc}~(4\%),
NA44~\cite{Bearden:2002ib}~(3.7\%),
NA49~\cite{Anticic:2004yj}~(5\%),
PHENIX~\cite{Adler:2003cb,Adcox:2003nr}~(5\%)
and PHOBOS~(15\%) experiments.
A remarkably smooth collision energy dependence of the `crossing' $p_T$
value is observed. At low energies, the abundance of produced pions is
naturally low compared to high collision energies, while baryon number
conservation ensures that a significant fraction of the large number of
initial state protons are found in the final state, thus the invariant
yields of protons and positive pions become comparable already at low
$p_T$. With increasing energy, this $p_T$ value grows, mainly due to the
approximately logarithmically increasing number of produced pions.
Our new result at 62.4 GeV fits smoothly into the trend established by 
earlier experiments.

\ack
%
%
%
%
This work was partially supported by U.S. DOE grants 
DE-AC02-98CH10886,
DE-FG02-93ER40802, 
DE-FG02-94ER40818,  
DE-FG02-94ER40865, 
DE-FG02-99ER41099, and
DE-AC02-06CH11357, by U.S. 
NSF grants 9603486, 
0072204,            
and 0245011,        
by Polish KBN grant 1-P03B-062-27(2004-2007),
by NSC of Taiwan Contract NSC 89-2112-M-008-024, and
by Hungarian OTKA grant (F 049823).



\end{document}